\RequirePackage{lineno}
\documentclass[twocolumn,10pt,nofootinbib]{revtex4}  
\usepackage{graphicx}  
\usepackage{dcolumn}   
\usepackage{bm}        
\usepackage{amssymb}   
\usepackage{multirow}
\usepackage{amsmath,amssymb,amsfonts}
\usepackage{float}
\usepackage{blindtext}
\usepackage{color}
\usepackage[toc,page]{appendix}
\usepackage{makecell}
\usepackage{hyperref}
\usepackage[caption=false]{subfig}
\usepackage{soul}
\usepackage{ulem}
\newcommand{\bef}{\begin{figure}}
\newcommand{\eef}{\end{figure}}
\newcommand{\bc}{\begin{center}}
\newcommand{\ec}{\end{center}}

\newcommand{\antinue}{\ensuremath{\overline{\nu}_{e}}}

\begin{document}
\title{\bf{Active-sterile neutrino mixing constraint using reactor 
antineutrinos with the ISMRAN setup}}
\author{S. P. Behera}
\email{shiba@barc.gov.in}
\author{D. K. Mishra}
\author{L. M. Pant}
\affiliation{Nuclear Physics Division, Bhabha Atomic Research Centre, Mumbai 
- 400085, India}
\begin{abstract}
In this work, we present an analysis of the sensitivity to the active-sterile 
neutrino mixing with the Indian Scintillator Matrix for Reactor Anti-Neutrino 
(ISMRAN) experimental setup at very short baseline. The 3 (active)$+$1 (sterile) 
neutrino oscillation model is considered to study the sensitivity of the 
active-sterile neutrino in the mass splitting and mixing angle plane. In this 
article, we have considered the measurement of electron antineutrino induced events 
employing a single detector which can be placed either at a single position or 
moved between a near and far positions from the given reactor core. Results 
extracted in the later case are independent of the theoretical prediction of 
the reactor antineutrino spectrum and detector related systematic uncertainties.
Our analysis shows that 
the results obtained from the measurement carried out at combination of the 
near and far detector positions are improved significantly at higher $\Delta 
m^{2}_{41}$ compared to the ones obtained with the measurement at a single detector 
position only. It is found that the best possible combination of near and far 
detector positions from a 100 MW$_{th}$ power DHRUVA research reactor core are 7 m 
and 9 m, respectively, for which ISMRAN setup can exclude in the range 1.4 
$eV^{2} \leq \Delta m^{2}_{41} \leq$ 4.0 $eV^{2}$ of reactor antineutrino anomaly 
region along with the present best-fit point of active-sterile neutrino 
oscillation parameters. At those combinations of detector positions, the ISMRAN 
setup can observe the active sterile neutrino oscillation with a 95$\%$ confidence 
level provided that $\sin^{2}2\theta_{14}\geq 0.09$ at $\Delta m^{2}_{41}$ = 1 
eV$^{2}$ for an exposure of 1 ton-yr. The active-sterile neutrino mixing sensitivity 
can be improved by about 22\% at the same exposure by placing the detector at 
near and far distances of 15 m and 17 m, respectively, from the compact proto-type 
fast breeder reactor (PFBR) facility which has a higher thermal power of 1250 
MW$_{th}$.
\end{abstract}

\maketitle

\section{INTRODUCTION}                                               %
\label{sec:intro}
Nuclear reactors are a copious source of electron antineutrinos due to beta decay of 
neutron-rich fission products. About 6 electron antineutrinos ($\antinue$) are 
produced per fission, corresponding to $\sim$10$^{20}$ antineutrinos per second 
from a reactor of thermal power 1 GW$_{th}$. Electron antineutrinos produced from the 
reactors played important roles in the history of particle physics, from 
establishing the existence of neutrinos~\cite{Reines:1953pu} to determining the 
non-zero value of mixing angle ($\theta_{13}$) by Double Chooz~\cite{Abe:2011fz}, 
Daya Bay~\cite{An:2012eh}, and RENO~\cite{Ahn:2012nd} experiments. These experiments 
used near and far detector(s) to cancel out correlated systematic uncertainties due 
to the reactor $\antinue$ flux and the dependence on the absolute flux 
and thus have significantly improved precision measurement of $\theta_{13}$ over 
single detector experiments. 

Observations of the reactor $\antinue$ flux suffer an anomalous and unexplained 
behavior. The theoretical calculation of the $\antinue$ flux by Mueller 
$et~al.$~\cite{Mueller:2011nm} and Huber ~\cite{Huber:2011wv} predicts 6$\%$ more 
events than those observed in several reactor experiments at small distances.
This is known as the ``reactor antineutrino anomaly" (RAA)~\cite{Mention:2011rk}. 
The source of this anomaly is not known yet. However, there are two possible proposed 
explanations for this discrepancy. One of them is an incomplete prediction of 
the antineutrino flux and energy spectrum from reactors, due to underestimated 
systematics of the measurements of beta spectra emitted after 
fission~\cite{Feilitzsch1982, Schreckenbach1985, Hahn1989} or of the conversion 
method~\cite{Mueller:2011nm,Huber:2011wv,Hayes2016,Huber2016}. The other 
explanation is the disappearance of $\antinue$s~while propagating from the source to 
detector due to active-sterile neutrino oscillations with mass squared difference 
$\sim$1 eV$^2$. In addition, there can be a third possible explanation 
represented by possible unknown processes that affect the measurements.

The measurement of the reactor \antinue~induced positron spectra shows a 
statistically significant excess of events over the prediction, particularly in the 
energy spectrum at the range of 5--7 MeV (the so-called reactor bump). It puts in 
question the correctness of the flux calculation or to explore explanations with a 
new physics. An excess of events in the \antinue~ spectra is observed by Double 
Chooz~\cite{Abe:2015rcp}, Daya Bay~\cite{An:2015nua}, and RENO~\cite{RENO:2015ksa} 
Collaborations as well as other short baseline experiments such as 
NEOS~\cite{Ko:2016owz}. The bump in energy spectra has been correlated to the 
power of the reactor~\cite{An:2015nua}, and may be due to the $^{235}$U 
fuel~\cite{An:2017osx}. To verify the hypothesis of the existence of active to 
sterile neutrino oscillation as a possible origin of the RAA, as well as to clarify 
the origin of the bump at 5 MeV in the \antinue~spectra at a very significant 
confidence level, several experiments are currently underway and will collect data 
soon~\cite{Boser:2019rta}.

To address the RAA, the short-baseline (SBL) experiments are aiming to measure the 
reactor antineutrino energy spectra at two or more different distances and are trying 
to reconstruct the \antinue s survival probability both as a function of energy and 
the source to detector distance, $L$. The measurement of data at two distances 
with respect to the measurement at a single position is less sensitive to the 
modification of the \antinue~ spectra due to the time evolution of fuel composition 
in the reactor core, known as the burn-up effect, which is a source of 
systematic uncertainty. The $L$ dependence is what gives the cleanest signal in the 
case of the sterile neutrino, and studying the ratio of the spectra measured at 
two different distances allows to avoid almost completely the problem of the 
theoretical spectrum. Based on this approach, several experiments have collected 
data to study the active-sterile neutrino oscillation. The DANSS 
collaboration~\cite{Alekseev:2018efk} has measured the positron energy spectra 
at 3 different distances from the reactor core. The distances were varied from 10.7 m 
to 12 m to observe the active-sterile neutrino oscillations. Their observation 
excludes a large fraction of RAA region in the sin$^2 2\theta_{14} - \Delta 
m^{2}_{41}$ plane and covers the parameter space up to sin$^2 2\theta_{14} <$ 0.01. 
The STEREO~\cite{Almazan:2018wln} collaboration has measured the antineutrino energy 
spectrum in six different detector cells covering baselines between 9 and 11 meters 
from the core of the ILL research reactor. The results based on the reactor ON data 
are compatible with the null active-sterile neutrino oscillation hypothesis and the 
best-fit of the RAA can be excluded at 97.5$\%$ confidence level (C.L.). The
PROSPECT collaboration has measured the reactor \antinue~ spectra using a movable 
segmented detector array. Their observation disfavors the RAA best-fit point at 
2.2$\sigma$ C.L. and constrains a significant portion of the previously 
allowed parameter space at 95$\%$ C.L.\cite{Ashenfelter:2018iov}. The Neutrino-4 
experiment has measured \antinue~ energy spectra by mounting the segmented detector 
on a movable platform which covers a baseline range from 6 to 12 meters. Their 
model-independent analysis excludes the RAA region at C.L. more than 3$\sigma$. 
However, the experiment has observed active-sterile neutrino oscillation at sin$^2 
2\theta_{14}$ = 0.39 and $\Delta m^{2}_{41}$ = 7.3 $eV^2$ at C.L. of 
2.8$\sigma$~\cite{Serebrov:2018vdw}. The Neutrino-4 best-fit is incompatible 
with PROSPECT bounds. To this end, the Indian Scintillator Matrix for Reactor 
Anti-Neutrino (ISMRAN) detector is proposed. It will be mounted on a movable trolley 
in order to place the complete setup at different distances with respect to the 
reactor core. Here we study its potential to observe active-sterile neutrino 
oscillation at SBL (L$<$ 25 m). An investigation is carried out employing the 
3~$+$~1 neutrino mixing model, where `3' refers to active neutrinos and `1' to 
sterile neutrinos. This is the only allowed active-sterile neutrino mixing 
scheme~\cite{Gariazzo:2017fdh} under the assumption of 4 neutrino model. The 
existence of active-sterile neutrino oscillation with mass squared difference $\Delta 
m_{41}^2 (= m_4^2 - m_1^2) \sim$1 eV$^2$ can be explored at the SBL experiment by 
measuring the reactor $\antinue$ flux which is reduced due to the fast active to 
sterile neutrino oscillation that is otherwise absent in the 3-neutrino mixing 
scheme. A similar study has been performed previously, considering a single 
detector which will be placed at a fixed distance from the reactor core by varying 
both reactor and detector related parameters~\cite{Behera:2019hfs}. To reduce 
the systematic uncertainties mentioned earlier, in this work we have considered 
various possible combinations of near and far positions for the same 1-ton detector 
which will be placed for a period of six months at each distance while constraining 
active-sterile neutrino oscillation parameters. It can be noted that, in case of 
a research reactor the burn-up period is small and the time evolution of fuel
 has less impact on the modification of \antinue~ spectra. Hence, we can either 
place it six months at each location or shift the position of the detector more 
frequently. On the other hand, a power reactor has a longer burn-up period. 
Therefore, it is important to consider the fuel evolution of the reactor with time, 
which can be minimized by changing the  position of the detector more frequently.

The article is organized in the following order. A detailed description of the 
ISMRAN setup and neutrino detection principle is discussed in 
Sec.~\ref{sec:detector} and Sec.~\ref{sec:detectionprinciple}, respectively. 
The phenomenon of active-sterile neutrino oscillation at SBL considering the `3+1' 
mixing model is described in Sec.~\ref{sec:osciprob}. The procedure for the 
incorporation of detector response on $\antinue$ induced simulated events is 
mentioned in Sec.~\ref{sec:simul}. In order to find out the ISMRAN setup 
sensitivity to the active-sterile neutrino oscillation parameters, a statistical 
method  on $\chi^{2}$ estimation considered in this study is discussed in 
Sec.~\ref{sec:chisq}. The sensitivity to active-sterile neutrino mixing at an 
exposure of 1 ton-yr is elaborated in Sec.~\ref{sec:results}. In 
Sec.~\ref{sec:summary}, we summarize our observations and discuss the implication of 
this work. 
\section{THE ISMRAN SETUP}\label{sec:detector} 
The one-ton active volume ISMRAN setup consists of 100 segmented plastic 
scintillator (PS) bars with a total volume of 1 m$^3$. The size of each PS bar is 100 
cm $\times$ 10 cm $\times$ 10 cm and is wrapped with aluminized mylar foils that have 
been coated with gadolinium. The gadolinium coating increases the detection 
efficiency of neutrons. At both ends, a PS bar is coupled with two 3 inch 
photo-multiplier tubes. More information on the detector and background 
measurements carried out at the experimental site can be found in 
Ref.~\cite{Mulmule:2018efw}. Due to the compact size of the detector, it can be 
easily maneuvered from one place to another. This is useful for the remote monitoring 
of the power of the reactor. The segmented detectors array can provide 
additional position information while reconstructing the neutrino induced events and 
thus will improve the active sterile neutrino mixing sensitivity of the ISMRAN 
setup. The energy and position information of an event will be extracted from the 
signals of the PS bars. All signals will be digitized with a CAEN-made digitizer. 
Details of the signal processing and data acquisition are given in 
Ref.~\cite{Mulmule:2018efw}. The ISMRAN setup active volume is surrounded with 
passive shielding material consisting of 10 cm thick Lead followed by 10 cm thick 
borated polyethylene in order to suppress both the natural and reactor related 
background such as gamma-rays and neutrons. Further, the setup will be surrounded 
by 1-inch thick scintillator plates for vetoing the cosmic muons.

The proposed ISMRAN setup will be placed at the DHRUVA research reactor facility in 
Bhabha Atomic Research Centre (BARC), India. The setup consists of inflammable 
plastic scintillator detectors, so they can be placed as close as possible to the 
reactor core. The closest possible distance at which the detector can be placed is 
about 7 m from the reactor core. The DHRUVA reactor core has a cylindrical shape with 
radius $\sim$1.5 m and height $\sim$3.03 m (defined as an extended 
source)~\cite{Agarwal:dhruva}. The reactor can operate at a maximum thermal power of 
100 MW$_{th}$ consuming natural uranium as fuel. In the future, it is planned to put 
the ISMRAN setup at other reactor facilities such as upgraded Apsra (U-Apsra) 
reactor~\cite{Singh:uapsara}, BARC, and, proto-type fast breeder reactor (PFBR), 
IGCAR, Kalpakkam, India~\cite{Chetal:pfbr}.

The U-Apsra reactor has a  compact core with a height of about 0.64 m 
and radius about 0.32 m which can operate at a maximum thermal power of 3 
MW$_{th}$~\cite{Singh:uapsara}. The closest possible distance at which the 
detector can be placed is about 4 m from the reactor core, which is an ideal position 
considering average $\antinue$ energy about 4 MeV and active-sterile neutrino 
oscillation at $\Delta m^2_{41}\simeq$ 1 eV$^2$. At this distance, $L/E$ value 
is of the order of 1 m/MeV. Hence, this will maximize the sensitivity to sterile 
neutrino masses at the eV-scale. On the other hand, PFBR is a relatively compact 
source with respect to DHRUVA reactor. The PFBR has dimensions of about 1 m in 
both radius and height. The PFBR can operate at a maximum thermal power of 1250 
MW$_{th}$ and employs mixed oxide (MOX, PuO$_{2}$-UO$_{2}$) as 
fuel~\cite{Chetal:pfbr}. The closest possible 
distance at which the detector can be placed is about 15 m from PFBR core.
These compact U-Apsra and PFBR reactors are ideal sources to utilize the ISMRAN 
setup for investigating the active-sterile neutrino mixing at a short 
distance. However, at such close distances, there are significant contributions 
from the reactor related background on the sterile neutrino sensitivity. At 
present, measurements of reactor related backgrounds are going on with a proto-type 
ISMRAN setup consists of 16 PS bars placed at a distance of 13 m from DHRUVA 
reactor core. The above-mentioned reactors are not only different in 
terms of their sizes and thermal power but also different with respect to their 
various fuel compositions as mentioned in Table~\ref{tab:reactortype}. These 
reactors have different fuel compositions and hence the measurements with ISMRAN 
setup will be different from existing worldwide experimental observations 
because of the different $\antinue$ fluxes at each reactor. This will be an ideal 
situation to compare with other results regarding the bump at 5 MeV. 
\begin{table*}
 \begin{center}
\caption{\label{tab:reactortype}{Reactor details}}
\begin{tabular}{ ccc}
    Reactors name & Thermal power(MW$_{th}$) & Fuel type \\
  \hline
 DHRUVA & 100.0 & Natural uranium  \\
 PFBR & 1250.0 & MOX(PuO$_{2}$-UO$_{2}$)  \\ 
 U-Apsra & 3.0 & U$_{3}$Si$_2$-Al (Low enriched $^{235}$U) \\
 \hline
\end{tabular}
\end{center}
\end{table*}
\section{THE \antinue~DETECTION PRINCIPLE}\label{sec:detectionprinciple}%
The PS bars in the ISMRAN setup act as a target as well as active detection material 
for the $\antinue$s. The basic principle of detection for $\antinue$s produced from 
the reactors is via the inverse beta decay (IBD) process. The IBD process is given 
by 
\begin{equation}
\label{eq:ibdreac} 
\bar\nu_{e} + p \rightarrow n + e^{+} .
\end{equation}
The minimum antineutrino energy required for the above reaction to occurs is 
about 1.80 MeV. In this process, the positron carries almost all of the 
available energy, loses it by ionization in the detector, and produces two 
$\gamma$-rays each having energy 0.511 MeV through annihilation process. This is 
the `prompt' signal. The neutron produced through the IBD process carries a 
few keV's of energy and gets thermalized within several $\mu$s in collisions with 
protons in the PS bar. The thermal neutron then gets captured by hydrogen (captured 
time of $\sim$ 200 $\mu$s) in the PS bar. This is the `delayed' signal. In 
this case, a monoenergetic gamma-ray of energy 2.2 MeV is produced, comparable 
to the gamma-ray energy originated from some of the natural backgrounds. To further 
increase the probability for neutron capture and improve the detection efficiency, PS 
bars are wrapped with gadolinium (Gd) coated aluminized mylar foil as both $^{155}$Gd 
and $^{157}$Gd have high thermal neutron capture cross-section. Hence, the 
reduced neutron capture time is $\sim$ 60 $\mu$s, observed in prototype ISMRAN 
setup~\cite{Mulmule:2018efw}. There is also a cascade of gamma-rays produced with 
a total energy of about 8 MeV due to neutrons captured in the gadolinium. Due 
to higher total energy, it is possible to distinguish these gamma-rays from the 
natural background. The coincidence of a prompt positron signal and a delayed signal 
from captured neutron uniquely identifies the IBD event. 

The detection of candidate events is dominated by two types of backgrounds. The first 
is the accidental background as a result of two random energy depositions in a time 
window corresponding to the captured time of the neutron. The other type of 
background is the correlated background originating from either spallation of cosmic 
muons, which produces fast neutrons, or fast neutrons coming from the reactor due to 
fission fragments. The prompt signal arises due to the energy loss of fast neutrons 
through scattering off protons and a delayed signal due to captured neutron in PS, 
both constitute a IBD-like event. However, efficient delayed coincidence 
technique allows us to suppress such types of backgrounds~\cite{SOguri}. These 
background contributions would affect the detector sensitivity, so it is essential to 
reduce them. This is discussed further in Sec.~\ref{sec:results}.
%
\section{NEUTRINO OSCILLATION PROBABILITY AT SHORT BASE LINE}%
\label{sec:osciprob}
There are three flavors of active neutrinos ($\nu_{e}$, $\nu_{\mu}$, $\nu_{\tau}$) in 
the standard model. Neutrinos are produced and detected as flavor states. However, 
they propagate as superpositions of mass eigenstates. The transformation 
between flavor and mass eigenstates is expressed by the 
Pontecorvo-Maki-Nakagawa-Sakata (PMNS)~\cite{Maki:1962mu} unitary matrix. The 
establishment of phenomena of neutrino oscillation and  measurements of the three 
generations of oscillation parameters are carried out by several 
experiments~\cite{GonzalezGarcia:2002dz,Kajita:2000mr,Bemporad:2001qy,deSalas:2017kay}.
At present, various experiments are aiming to measure the oscillation
parameters more precisely. However, beyond these three active neutrinos, world-wide 
research programs are underway and some experiments will take data in the near 
future to explore the possible existence of active-sterile neutrino oscillation. The 
active-sterile neutrino mixing sensitivity of the ISMRAN setup is studied 
considering the `3+1' neutrino mixing model which was mentioned earlier. In this 
model, the 3 generation PMNS matrix are expanded to the 3$+$1 generation, where 
``3" stands for three active neutrinos and ``1" for a sterile neutrino~($\nu_{s}$). 
The order of rotation and elements of the mixing matrix are given in 
Ref.~\cite{Behera:2019hfs}. At a small value of mixing angle $\theta_{14}$ and source 
to detector distance of few meters ($<$ 100m), the 3$+$1 oscillation scheme can be 
simplified to a two neutrino scheme and the $\nu$ survival probability is 
approximated to  
\begin{equation}
\label{eq:prob}
P_{\nu_e\nu_e}(E_{\nu},L) \simeq 1-\sin^22\theta_{14}\sin^2\left(\frac{1.27 
\Delta m^2_{41}{L}}{E_{\nu}}\right),
\end{equation}
where $E_{\nu}$ is the neutrino energy (in MeV), $L$ is the distance (in m) between 
the production and the detection of the neutrino and $\Delta m^2_{41}$ is the squared 
masses difference (in eV$^2$) between the two neutrino mass eigenstates. The 
oscillation parameters $\Delta m^2_{41}$ and $\sin^22\theta_{14}$ are given by
\begin{equation}
\Delta m^2_{41}  =  m^{2}_{4}- m^{2}_{1}~;~~
\sin^22\theta_{14}  =  4 |U_{e4}|^2(1 - |U_{e4}|^2), 
\end{equation}
where $U_{e4}$ is an element of the unitary mixing matrix. 
The oscillation probabilities for antineutrinos can be obtained by replacing the 
mixing matrix elements $U$s with their complex conjugate ($U^{\ast}$s). However, at 
SBL experiments, the oscillation probability is independent of the $CP$-violating 
phases~\cite{Palazzo:2013bsa}. Hence the oscillation probability given in 
Eq.~\ref{eq:prob} is the same for antineutrino. Experimental studies on 
neutrino oscillations aim to determine the mass parameter $\Delta m^2_{41}$ and the 
mixing angle $\sin^22\theta_{14}$. These parameters can be obtained by measuring the 
neutrino flux at different energies and distances. The present best-fit values of 
active-sterile neutrino oscillation parameters are $\Delta m^2_{41} \simeq $ 1.30 
eV$^{2}$ and $\sin^22\theta_{14} \simeq$ 0.049 ~\cite{Gariazzo:2018mwd} extracted 
from the combined analysis of data taken by NEOS and DANSS collaborations. Similar 
values are also found from global analysis~\cite{Dentler:2018sju}. At these values of 
the neutrino oscillation parameters, the possible existence of sterile neutrino at 
SBL experiments can be observed by looking at the distortions of the $\antinue$ 
energy spectrum at short distances which are otherwise absent in the three active 
neutrino oscillations. However, these distortions are smeared out for longer source 
to detector distances and the phase factor of the oscillation probability averaged 
out to 1/2. This leads to the survival probability of $1 -\sin^2 2\theta_{14}/ 2$. 
Hence, we lose the information regarding $\Delta m^2_{41}$ and can measure only the 
mixing angle $\theta_{14}$. However, measuring the oscillation parameters by 
measuring $\antinue$s with a detector placed at only one distance from the reactor 
core and comparing it with the prediction is not enough, since the theoretical 
calculation of the $\antinue$ energy distribution is not reliable enough. Therefore, 
the most reliable way to observe such distortions is to measure the $\antinue$ 
spectrum with the same detector at various distances. In this case, the shape and 
normalization of the $\antinue$ spectrum as well as the detector efficiency are 
canceled out. Alternatively, one can put two same types of detectors at near and 
far positions in order to avoid the assumption of constant reactor flux. In such a 
case, although two detectors of the same type, their response and other detector 
related parameters may not be the same, which will introduce the detector related 
uncertainties.
\begin{table*}[ht]
 \begin{center}
\caption{\label{tab:para}{Fractional contributions of each element 
to the reactor thermal power and the parameters used to fit the neutrino spectrum}}
\begin{tabular}{ c|ccc|cccccc}
    Element & \centering{a} &&& $b_0$ & $b_1$ & $b_2$ & $b_3$ & $b_4$& $b_5$\\
\multicolumn{2}{c}{}\\
& \text{U-Apsra} & \text{DHRUVA} & \text{PFBR} \\
  \hline
 $^{235}$U & 0.90 & 0.58 & 0.0093& 4.367 & -4.577 & 2.1 & -0.5294 & 0.06186 
 & -0.002777 \\

 $^{239}$Pu & 0.07 & 0.30 & 0.71 & 4.757 & -5.392 & 2.63 & -0.6596 & 0.0782 
 & -0.003536 \\ 

 $^{241}$Pu & 0.01 & 0.05 & 0.11 & 2.99 & -2.882 & 1.278 & -0.3343 & 0.03905 
 & -0.001754 \\

 $^{238}$U & 0.02 & 0.07 &  0.10 & 0.4833 & 0.1927 & -0.1283 & -0.006762 & 0.002233 
 & -0.0001536 \\
  \hline
\end{tabular}
\end{center}
\end{table*}
\section{SIMULATION PROCEDURE}
\label{sec:simul}              %
The potential of the ISMRAN setup on finding active-sterile neutrino oscillation 
sensitivity will be explored by using antineutrinos produced from various types of 
reactor facilities such as the U-Apsra, DHRUVA, and PFBR. The number of $\antinue$s~ 
produced from the reactor not only depends on the thermal power but also on their 
fuel compositions. The energy spectrum of the $\antinue$s produced from the reactor 
is different for different isotopes. The parameterization for $\antinue$ flux assumed 
in the present analysis is as 
follows:
\begin{equation}
f(E_{\antinue}) = \sum_{i =~0}^{4} a_{i} \exp\bigg(\sum_{j =~0}^{6} b_{j} 
E_{\antinue}^{j-1}\bigg),
\label{nuflux}
\end{equation}
where `$a_i$' is the fractional contribution from the $i$-th isotope to the reactor 
thermal power,`$b_j$'s are the constant terms used to fit the antineutrino energy 
spectra, and $E_{\antinue}$ is neutrino energy in MeV. The fractional contributions 
of each isotope to the reactor thermal power and the parameter lists used to fit the 
neutrino energy spectra are summarized in Table~\ref{tab:para}. Both $a_i$ and $b_j$ 
values for various isotopes are taken from Ref.\cite{Zhan:2008id} and 
Ref.~\cite{Huber:2016fkt} for the DHRUVA and PFBR reactors, respectively. In the case 
of the U-Apsra reactor, we have assumed the fractional contributions of each 
isotope to the reactor thermal power as mentioned in Table~\ref{tab:para}. The 
list of parameters used to fit the $\antinue$ spectra due to $^{235}$U, $^{239}$Pu 
and $^{241}$Pu are considered from Ref.~\cite{Huber:2011wv} and for $^{238}$U is 
taken from Ref.~\cite{Mueller:2011nm}. We have also considered the spatial 
variation of  $\antinue$ flux due to a finite size cylindrical reactor that depends on 
the radius and height of the core which is given by~\cite{neu_flux}, 
\begin{equation}
\phi = \phi_{0}~J_{0}(2.405 r/R)~cos(\pi z/H)
\end{equation}   
where $\phi_0$ represents the flux at the center of the reactor core, $R$ is 
the radius of the cylindrical reactor core, $H$ is the height, $J_0$ is the 
zeroth-order Bessel function of the first kind where $r$ ($0\leq r\leq R$) and z ($ 0 
\leq z \leq H$) are the vertex positions of the $\antinue$s produced in the reactor. 
The interaction cross-section of $\antinue$ for the inverse IBD process is 
given by~\cite{Vogel:1999zy}
\begin{equation}
\sigma_{IBD} = 0.0952 \times 10^{-42} \mathrm{cm}^2 (E_{e^{+}} p_{e^{+}} /\mathrm{MeV}^2),
\end{equation}
where $E_{e^{+}}$ = $E_{\antinue} - (m_n -m_p)$ is the positron energy, 
neglecting the recoil neutron kinetic energy,  and $p_{e^{+}}$ is the positron 
momentum. The detector resolution is folded on the true positron (kinetic) energy 
spectrum by assuming a standard Gaussian form of the energy resolution:

\begin{equation}
R(E_{e^{+}},E_{e^{+},T}) = 
\frac{1}{{\sqrt{2\pi}}\sigma} \exp(-\frac{(E_{e^{+}} - E_{e^{+},T})^2}{2\sigma^2})\,
\label{Ereso}
\end{equation}
where $E_{e^{+},T}$ and $E_{e^{+}}$ are the simulated true and  observed positron 
energy, respectively. The detector resolution considered for this study is in the 
form $\sigma/E_{e^{+}} ~\sim$ 20$\%$/$\sqrt{E_{e^{+}}}$. The neutrino induced 
events are distributed in terms of positron energy spectrum. There are a total of 80 
bins in the $e^+$ energy range of 0--8 MeV that are considered. The number of events 
in $i$-th energy bin after incorporating the detector resolution is given as
\begin{equation}
N_{i}^{r} = \sum_{k} K_{i}^{k}(E_{e^{+},T}^{k}) n_{k}
\label{nevt}
\end{equation}
The index $i$ corresponds to the measured energy bin, $N_{i}^{r}$ corresponds to the 
number of reconstructed events, $k$ is summed over the true energy of positron and 
$n_{k}$ is the number of events in $k$-th true energy bin. Further, $K_{i}^{k}$ is 
the integral of the detector resolution function over the $E_{e^{+}}$ bins and is 
given by

\begin{equation}
K_{i}^{k} = \int_{E_{e^{+},L_{i}}}^{E_{e^{+},H_{i}}} R(E_{e^{+}},E_{e^{+},T})~ 
dE_{e^{+}}
\end{equation}
The integration is performed between the lower and upper boundaries of the measured 
energy ($E_{e^{+},L_{i}}$ and $E_{e^{+},H_{i}}$) bins. In the present analysis, we 
have assumed 25$\%$ detection efficiency, 80$\%$ fiducial volume of the detector, and 
70$\%$ reactor duty cycle for an exposure of 1 ton-year. Both the production point of 
neutrinos in the reactor core and the interaction point in the detector are generated 
using a Monte-Carlo method.

\section{SENSITIVITY ESTIMATION}
\label{sec:chisq}  
The active-sterile neutrino mixing sensitivity of an experiment can be extracted by 
two independent pieces of information. The first is by knowing the $\antinue$ energy 
spectrum, flux, and cross-section accurately. From this, the total number of 
$\antinue$ induced events expected within the detector can be estimated for a given 
oscillation hypothesis and compared with the measured one. This is known as a ``rate 
only" analysis. The second case is a relative change of event rate as a function of 
the source to detector distance and $\antinue$ energy that can be compared with the 
predictions taking different oscillation hypotheses, without constraining the 
integral number of events. Using this method to find the sensitivity of the 
oscillation parameters is known as ``shape only" analysis. A combination of rate 
only and shape only analyses are used (known as ``rate + shape" analysis) in order 
to maximize the experimental sensitivity. These methods are affected by different 
systematic uncertainties. 

A statistical analysis of simulated event distribution for an exposure of 1 ton-year 
is performed in order to quantify the sensitivity of ISMRAN setup to the 
active-sterile neutrino mixing parameters $\theta_{14}$ and $\Delta m^2_{41}$. The 
detector response is incorporated in both theoretically predicted (events without 
active-sterile neutrino oscillation) and number of events expected due to 
active-sterile neutrino oscillation. The exclusion limit is extracted by estimating 
the $\chi^2$ for each value of $\Delta m_{41}^2$ with scanning over the various 
values of $\sin^2 2\theta_{14}$, and determining the boundary of the corresponding 
$\chi^2$ [e.g. $\chi^2$ = 5.99 for 95\% confidence limit(C.L.)]. Based on the ``rate 
+ shape" analysis, the definition of $\chi^{2}$ is taken from Ref.~\cite{pu} and is 
given by
\begin{equation}
\chi^{2} =\sum_{n=0}^{N} \bigg(\frac{ 
N_{n}^{th}-N_{n}^{ex}}{\sigma(N_{n}^{ex})}\bigg)^{2} + \sum_{i=0}^{k} \xi_{i}^{2}
\label{eq:chi1}
\end{equation}%
where $n$ is the number of energy bins, $N_{n}^{ex}$ is the expected number of 
observed events (with oscillations), and $N_{n}^{th}$ is the number of theoretically 
predicted events (without oscillations). The theoretically predicted events, 
$N_{n}^{th}$ are calculated considering reactor antineutrino flux as given by the 
Huber and Mueller model mentioned in Eq.~\ref{nuflux}, the IBD cross section, the 
detection efficiency, and detector energy resolutions. The simulated oscillated 
event, $N_{n}^{ex}$ is estimated by folding the oscillation probability on 
$N_{n}^{th}$. $N_{n}^{th}$ carries the information about systematic uncertainties 
given by

\begin{equation}
N_{n}^{th} = N_{n}^{'th}\bigg(1+\sum_{i=0}^{k}\pi_{n}^{i}\xi_{i}
\bigg)+\mathcal{O}(\xi^{2})
\label{eq:chi2}
\end{equation}
where $N_{n}^{'th}$  is the theoretically predicted event spectrum given by 
Eq.~\ref{nevt}. In the above $\pi_{n}^{i}$ is the strength of the coupling between 
the pull variable $\xi_{i}$ and $N_{n}^{'th}$. The $\chi^2$ is minimized with respect 
to pull variables {$\xi_i$}. The index $i$ in Eqs.~\ref{eq:chi1} and ~\ref{eq:chi2} 
runs from 0 to k, where $k$ is the total number of systematic uncertainties. We have 
considered four systematic uncertainties in our analysis. These include 3$\%$ 
normalization uncertainty (including reactor total neutrino flux, number of target 
protons, and detector efficiency), a nonlinear energy response of the detector by 
1$\%$, and, uncertainty in the energy calibration by 0.5$\%$. The uncorrelated 
experimental bin-to-bin systematic error of 2$\%$, which could 
occur due to insufficient knowledge of a source of 
background~\cite{Huber:2003pm}, is also considered. 
The definition given in Eq.~\ref{eq:chi1} includes both the rate and spectral shape 
information of neutrino induced events.

In the case of `rate only' analysis, the $\chi^{2}$ is estimated by 
integrating over energies as a single bin and setting all the systematic 
uncertainties to zero except the normalization uncertainty. It can be noted 
that the rate only analysis is sensitive to the active-sterile neutrino mixing angle. 
The `shape only' analysis is carried out considering the spectral shape information 
by setting the penalty term due to total reactor neutrino flux to zero. In 
this method, the oscillation frequency of $\Delta m^2_{41}$ from the energy dependent 
disappearance of the reactor $\antinue$ is considered without using the information 
on the total-rate deficit. In the present analysis, we have studied the sensitivity 
of the detector considering `shape only', `rate only' as well as combined `rate + 
shape' analysis separately, and compared the results obtained from each method.
\begin{figure*}[t]
\includegraphics[width=1.0\textwidth]{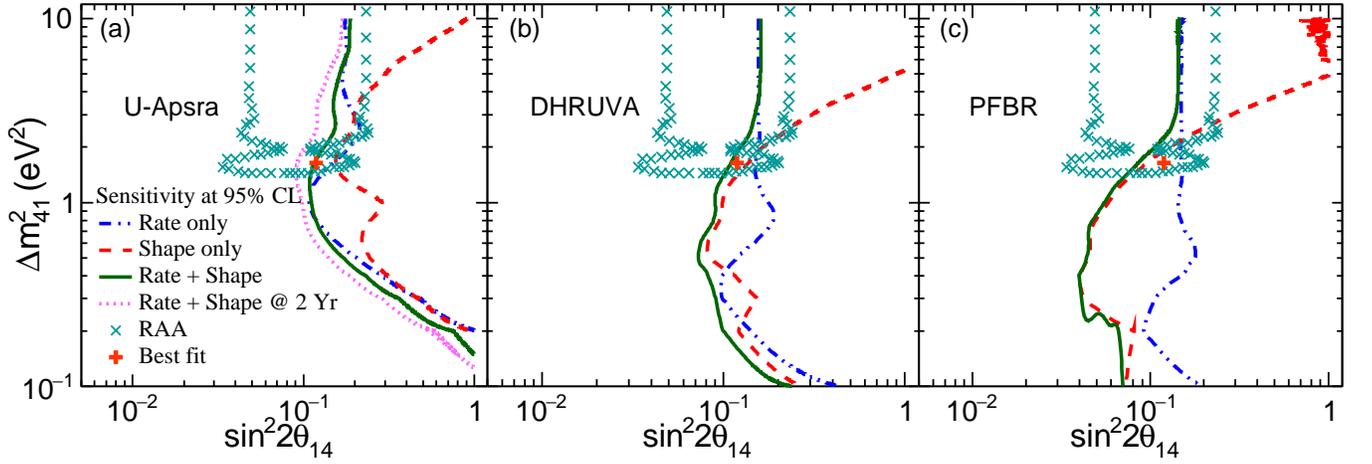}
\caption{ \label{fig:diffreactors} The expected active-sterile neutrino mixing 
sensitivity of ISMRAN setup in the sin$^{2}\theta_{14} - \Delta m^{2}_{41}$ 
plane. The left panel shows when the detector is placed at 4 m from the 
U-Apsra reactor core, the middle panel shows the case where the detector is  
positioned at 13 m from DHRUVA reactor core, and the right panel represents the 
study for which detector is placed at distance of 20 m from PFBR reactor core.} 
\end{figure*}

Since there is $\sim$~6$\%$ uncertainty in the theoretical prediction of reactor 
neutrino flux, it is essential either to build two identical detectors and 
locate one at near site and the other at far site or a single detector 
placed for some previously established time periods at 
the near and far position for certain periods in order to measure the 
active-sterile neutrino oscillation parameters precisely. However, we have 
considered various possible combinations near and far positions of the same detector 
to reduce the systematic uncertainties. For the two detectors case, the chi-square is 
defined as follows~\cite{Seo:2016uom},
\begin{equation}
\chi^{2} =\sum_{n=0}^{N} \bigg(\frac{ 
O_{n}^{F/N}-T_{n}^{F/N}}{\sigma(O_{n}^{F/N})}\bigg)^{2}, 
\label{eq:chiratio}
\end{equation}%
where $O_{n}^{F/N}$  is the simulated far-to-near ratio of oscillated events in 
$n$-th energy bin, $T_{n}^{F/N}$ is the expected far-to-near ratio of without 
oscillated events, and $\sigma(O_{n}^{F/N})$ is the statistical uncertainty of the 
oscillated event ratio $O_{n}^{F/N}$. It can be noted here that we have only 
considered the event spectra which will be measured at different far to near 
distances. The above definition of $\chi^2$ does not depend on the exact knowledge 
of the reactor power, absolute $\antinue$ flux, burn up effects, and detector related 
uncertainties. The definition of chi-square given in Eq.~\ref{eq:chiratio} is 
modified while considering the background for both the far and near detectors which 
is as follows,
\begin{equation}
\chi^{2}_{bkg} =\sum_{n=0}^{N} \bigg(\frac{ 
O_{n}^{F/N}-T_{n}^{'F/N}}{\sigma(O_{n}^{F/N})}\bigg)^{2} + \sum_{d={N,F}} \xi_{d}^{2} , 
\label{eq:redchiratio}
\end{equation}%
where $T_{n}^{'F/N}$ is defined as
\begin{equation}
T_{n}^{'F/N}=T_{n}^{F/N}\bigg(1+\sum_{d={N,F}}\pi_{n}^{d}\xi_{d} .
\bigg)+\mathcal{O}(\xi^{2})
\label{eq:chi3}
\end{equation}
In Eq.~\ref{eq:chi3}, $\pi_{n}^{d}$ is the strength of the coupling between the pull 
variable $\xi_{d}$ and $T_{n}^{F/N}$. The index $d$ in Eqs.~\ref{eq:redchiratio} and 
~\ref{eq:chi3} is for the near and far detectors. The background uncertainty is 
assumed to be 10.0$\%$ and 6.0$\%$ for near and far detectors, respectively.
\section{RESULTS AND DISCUSSIONS}
\label{sec:results}  
A study on active-sterile neutrino mixing sensitivity has been performed previously 
with the ISMRAN setup placed at a fixed distance from the reactor core while varying 
both the reactor and detector related parameters~\cite{Behera:2019hfs}. In the 
present study, we have considered the system in which the measurement will be carried 
out by placing the same detector at multiple positions with respect to the reactor 
core in order to cancel out the systematic uncertainties. The detector 
sensitivities to active-sterile oscillation parameters are compared by measuring the 
$\antinue$s produced from various types of reactors which are mentioned in 
Table~\ref{tab:reactortype}. 
\begin{figure*}[ht]
\begin{center}
\includegraphics[width=1.0\linewidth]{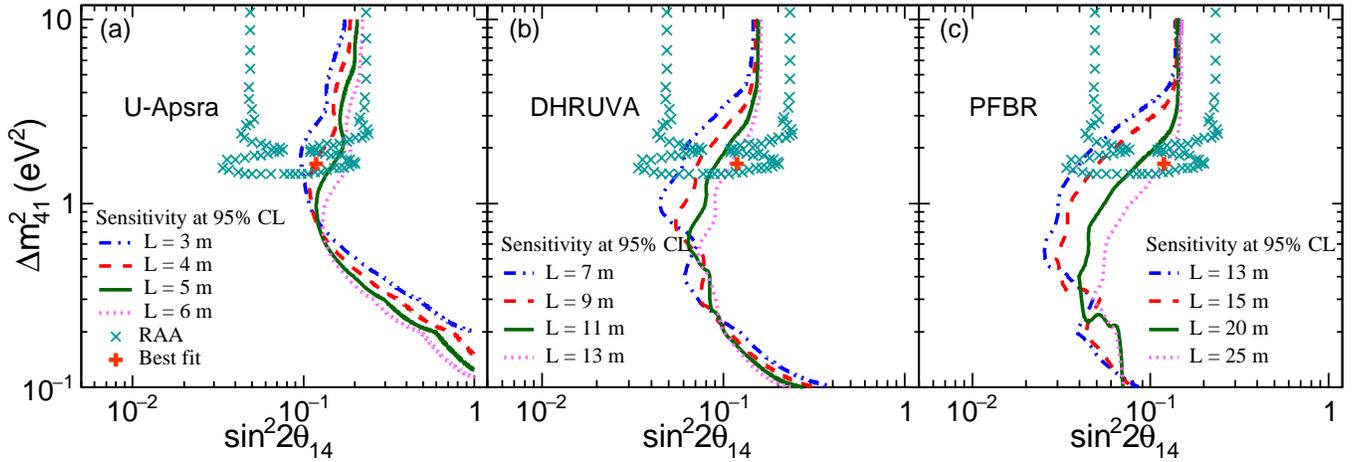}
\caption{ \label{fig:allreactordist} The comparison of the expected active-sterile 
neutrino mixing sensitivity for the ISMRAN setup placed at different source to 
detector path lengths for the U-Apsra (left panel), DHRUVA (middle panel), and PFBR 
(right panel) reactors.} 
\end{center}
\end{figure*}

\begin{figure*}[ht]
\begin{center}
\includegraphics[width=0.8\linewidth]{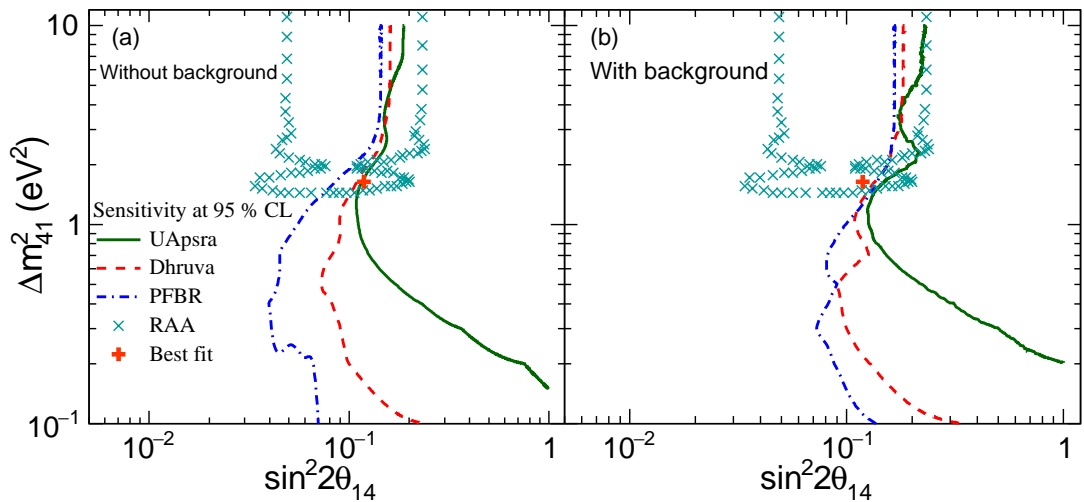}
\caption{ \label{fig:allreactorbkg} The comparison of active-sterile neutrino 
mixing sensitivity for the ISMRAN setup placed at fixed distances of 4 m, 13 m, and 
20 m from the U-Apsra, DHRUVA and PFBR reactors, respectively. The left (right) 
panel is without (with) inclusion of background in the simulated events. A signal
to background ratio of 1 is considered.}
\end{center}
\end{figure*}
\subsection{Detector at fixed distance}
Figure~\ref{fig:diffreactors} shows the active-sterile neutrino oscillation 
sensitivity of ISMRAN setup in the sin$^22\theta_{14}$ -- $\Delta m^2_{41}$ plane at 
95$\%$ C.L. for an exposure of 1 ton-yr. The left, middle, and right panels 
represent the results by placing the single detector at 4 m, 13 m, and, 20 m 
distances from the U-Apsra, DHRUVA, and PBFR reactor cores, respectively. The 
dashed-dotted blue, dashed red, and solid green lines, respectively, show the 
sensitivity by performing the `rate only', shape only', and a combination of `rate + 
shape' analysis. 

In the case of the ISMRAN setup at the U-Apsra reactor facility, the shape of the 
sensitivity curve at low $\Delta m^2_{41}$ ($e.g.$ 0.1 $\lesssim \Delta m^2_{41}$ 
(eV$^2$) $\lesssim$ 0.6) region shows a linear dependence between sin$^2 
2\theta_{14}$ and $\Delta m^2_{41}$ in a logarithmic scale. This is because the 
typical neutrino oscillation lengths are much larger compared to the size of the 
detector. Hence, the $\antinue$ survival probability mentioned in Eq.~\ref{eq:prob} 
approximates to $P_{\antinue\antinue}(E_{\antinue},L) \approx 1 - C\sin^2 
2\theta_{14} \times \left(\Delta m^2_{41}\right)^2$, where $C$ is a constant. It is 
observed that the `shape only' analysis has poor sensitivity to the oscillation 
parameters in the range 0.3 $< \Delta m^2_{41} (eV^2) <$ 1.5 and for $\Delta 
m^2_{41}>$ 3.0 $eV^2$ as compared to both `rate only' and `rate + shape' analysis. In 
the lower $\Delta m^2_{41}$ region, the shapes of the flux distributions are poorly 
affected by the oscillation deformations, as oscillations do not have enough space to 
fully develop. In the higher $\Delta m^2_{41}$ region, systematic uncertainties due 
to the antineutrino source dominate over statistical uncertainties. Also at higher 
$\Delta m^2_{41}$, the high-frequency oscillation probability gets averaged out 
due to the detector energy resolution. Both factors mentioned above result a 
gradual decrease of the shape discriminating power. In the parameter range $\Delta 
m^2_{41} \sim$ 0.6 - 1.3 $eV^2$, the ISMRAN setup has a maximum sensitivity with 
`rate only' and rate + shape analysis shown in left-panel of 
Fig.~\ref{fig:diffreactors}. It is found that results from both `rate only' and `rate 
+ shape' analysis overlap for $\Delta m^2_{41} \geq 5.0~ eV^2$. In this regime, the 
oscillation frequencies are large, and oscillations are suppressed by the detector 
energy resolutions and distribution of antineutrino path lengths. In case of 
`rate + shape' analysis, the rate deficit can be used to infer the $\sin^2 
2\theta_{14}$ mixing parameter, leading to contours that do not depend on the 
squared mass splitting $\Delta m^2_{41}$. It has been concluded from the above 
study that `rate + shape' analysis procedure has the best sensitivity to the 
oscillation parameters as compared to both `rate only' and `shape only' analyses. 
The dotted magenta line shows the sensitivity due to `rate + shape' analysis for an 
exposure of 2 ton-yr. It shows an overall improvement in active-sterile 
neutrino mixing sensitivity due to the increase in statistics.

A similar behavior has been observed in the active-sterile neutrino mixing 
sensitivity of the detector using various analysis methods as mentioned above
by considering the neutrino produced from DHRUVA and PFBR reactors. From this study,
it is observed that the ISMRAN setup can exclude a small portion of RAA using the `rate 
+ shape' analysis. In the above study, estimation of the sensitivity of ISMRAN is 
done by considering various reactor core sizes as well as path lengths from source to 
detector. It is found that the detector has the best sensitivity to the
oscillation parameters for a compact core compared to an extended one, due to the
large uncertainty in path lengths in the later. It can be noted here that rest of
our study has been carried out considering `rate + shape' analysis method.
\begin{figure*}[t!]
 
\includegraphics[width=1.0\linewidth]{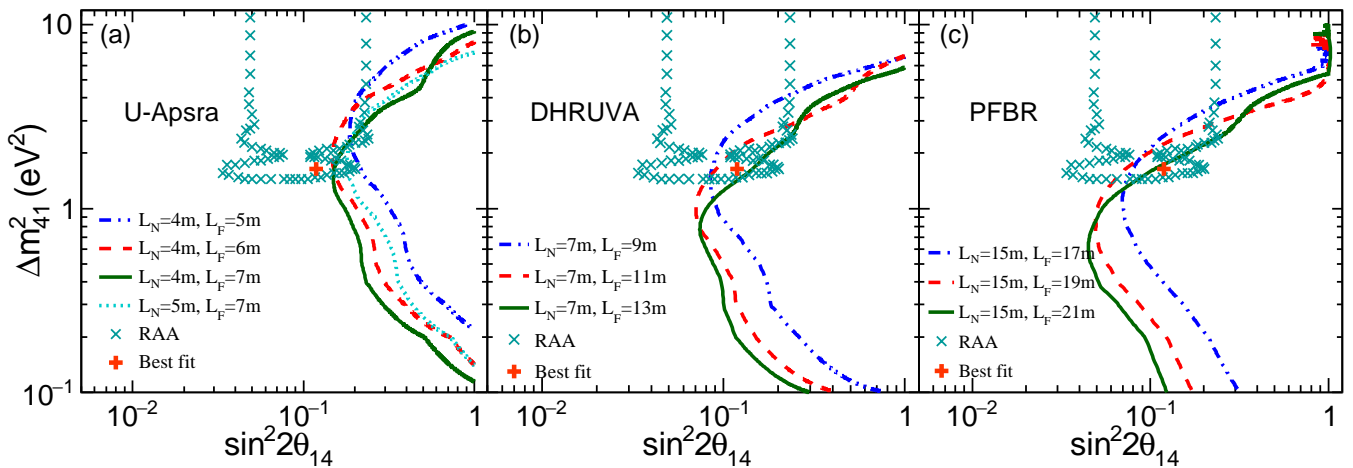}
\caption{ \label{fig:NFDet} The expected active-sterile neutrino mixing 
sensitivity of the ISMRAN setup in the sin$^{2}2\theta_{14} - \Delta m^{2}_{41}$ 
plane with a single detector which will be placed at a combination of near and 
far distances from the different reactor cores. The left, middle and, right 
panels represent results due to the U-Apsra, DHRUVA, and PFBR reactor facilities, 
respectively. } 
\end{figure*}

Figure~\ref{fig:allreactordist} shows the active-sterile neutrino sensitivity of 
the detector at various distances from the reactor cores to the center of the 
detector. In our calculation, both antineutrino vertices and their interaction in 
the detector are generated randomly using the MC method which was mentioned earlier.
At lower $\Delta m^2_{41}$ ($\simeq 0.1~eV^2$), the 
detector has the best sensitivity to the oscillation parameters by carrying out
the measurements at the PFBR facility, due to high thermal power and relatively
compact core. It is observed that 
the detector sensitivity improves with reducing the distance for higher $\Delta 
m^2_{41}$ ($> 1.0~eV^2$). By reducing the distance from the 
reactor, the event statistics are increased and hence the experimental sensitivity.
However, it is important to consider other shielding material structures 
surrounding the reactor core and associated reactor backgrounds while moving closer 
to the source.

Furthermore, the active-sterile neutrino sensitivity of the detector has been 
studied with and without the inclusion of background 
for three different types of reactors. Figure~\ref{fig:allreactorbkg}(a) shows the 
detector sensitivity in the sin$^{2}\theta_{14} - \Delta m^{2}_{41}$ plane without 
inclusion of background at an exposure of 1 ton-yr. The maximum sensitivity at lower 
$\Delta m^2_{41}$ ($< 1.0~eV^2$) is observed when measurements are done at the PFBR 
reactor facility. In the mass region of 1.5 $\leq \Delta m^2_{41} (eV^2) \leq$ 6.0, 
sensitivities are comparable for all the reactors. Figure~\ref{fig:allreactorbkg}(b) 
shows the detector sensitivity with the inclusion of background assuming a signal (S) 
to background (B) ratio of 1. A combination of backgrounds is 
considered~\cite{Heeger:2012tc}, such as a 1/$E^2$ dependence that represents 
the spectral shape due to accidental backgrounds that arises from intrinsic detector 
radioactivity and a flat distribution in antineutrino energy due to contributions 
from fast neutrons. We have considered an associated 10$\%$ systematic uncertainty 
due to these backgrounds. It is observed that with the contribution of background, 
the active-sterile neutrino mixing angle sensitivity of the detector is further 
reduced. However, a small portion of the RAA region can be excluded using all the 
available neutrino sources, but the best-fit point as well as the
remaining region of the RAA can be excluded with higher statistics, good detector 
energy resolution and with an improved signal to background ratio.
\subsection{Detector at combination of distances}
The discussions in the previous sub-section are based on a single detector placed at 
a fixed distance from the reactors. However, the systematic uncertainties due to 
reactor neutrino flux, as well as the detector play a major role when 
determining the active-sterile
neutrino mixing sensitivity. In order to reduce the systematic uncertainties, we have 
considered combinations of near and far positions for the same detector from the 
reactor core for periods of six/twelve months at each location. For example, the 
detector can be placed at a near distance of 7 m from the DHRUVA reactor core 
for a period of six months and then at a far distance of 13 m for six months, for 
 a total exposure of 1 ton-yr. 

Figure~\ref{fig:NFDet} shows the ISMRAN setup sensitivity to 
active-sterile neutrino oscillation parameters in the $\sin^2 2\theta_{14} - 
\Delta m^2_{41}$ plane at 95\% C.L. using several combinations of near and far 
detector positions from various reactor facilities, with an exposure of 1 
ton-yr. A small portion of the RAA region can be excluded using \antinue~s 
produced from three reactors. At lower $\Delta m^2_{41}$ ($< 1.0~eV^2$), the 
detector sensitivity to the oscillation parameters is the best for the 
measurements carried out at the PFBR reactor. In the mass region of 1.0 $\leq \Delta 
m^2_{41} (eV^2) \leq$ 10.0, the detector can have the best sensitivity using 
\antinue~s from U-Apsra reactor. \footnote{At a given near detector position, 
measurement of active-sterile neutrino oscillation parameters could be improved 
further by increasing the size of the far detector as mentioned in 
Ref.~\cite{Heeger:2012tc}. However, in the present study we have not considered the 
latter option.} It can be noted that a power reactor has a longer burn-up period (more than 1 year). So there is  a change in the reactor $\antinue$ flux with 
time. This can lead to reduction of the detector sensitivity to the 
active-sterile oscillation parameters if one is placing a single detector 6 months each 
at near and far position from the reactor core. In such a case, it is better to
frequently change the position of the single detector and obtain a better sensitivity.

Further study has been performed in order to find out the effect of background 
considering the near and far sites of 7 m and 9 m from the DHRUVA reactor core. 
Similar types of background are considered as mentioned earlier for a single 
detector. In this case, we have assumed two different cases of background, S/B = 1 
and 2. Figure ~\ref{fig:bkgtwodet} shows the comparison of 
the detector sensitivity between the ideal case and two scenarios with
different $S/B$ values. Due to the inclusion of
backgrounds, the sensitivity of the detectors is reduced in the entire considered
$\Delta m^2_{41}$ range and a substantial reduction is observed for $\Delta 
m^2_{41}<1.0$ eV$^2$. There is a small reduction of sensitivity in the RAA 
region by including the backgrounds.
\bef[ht!]
\bc
\includegraphics[width=0.47\textwidth]{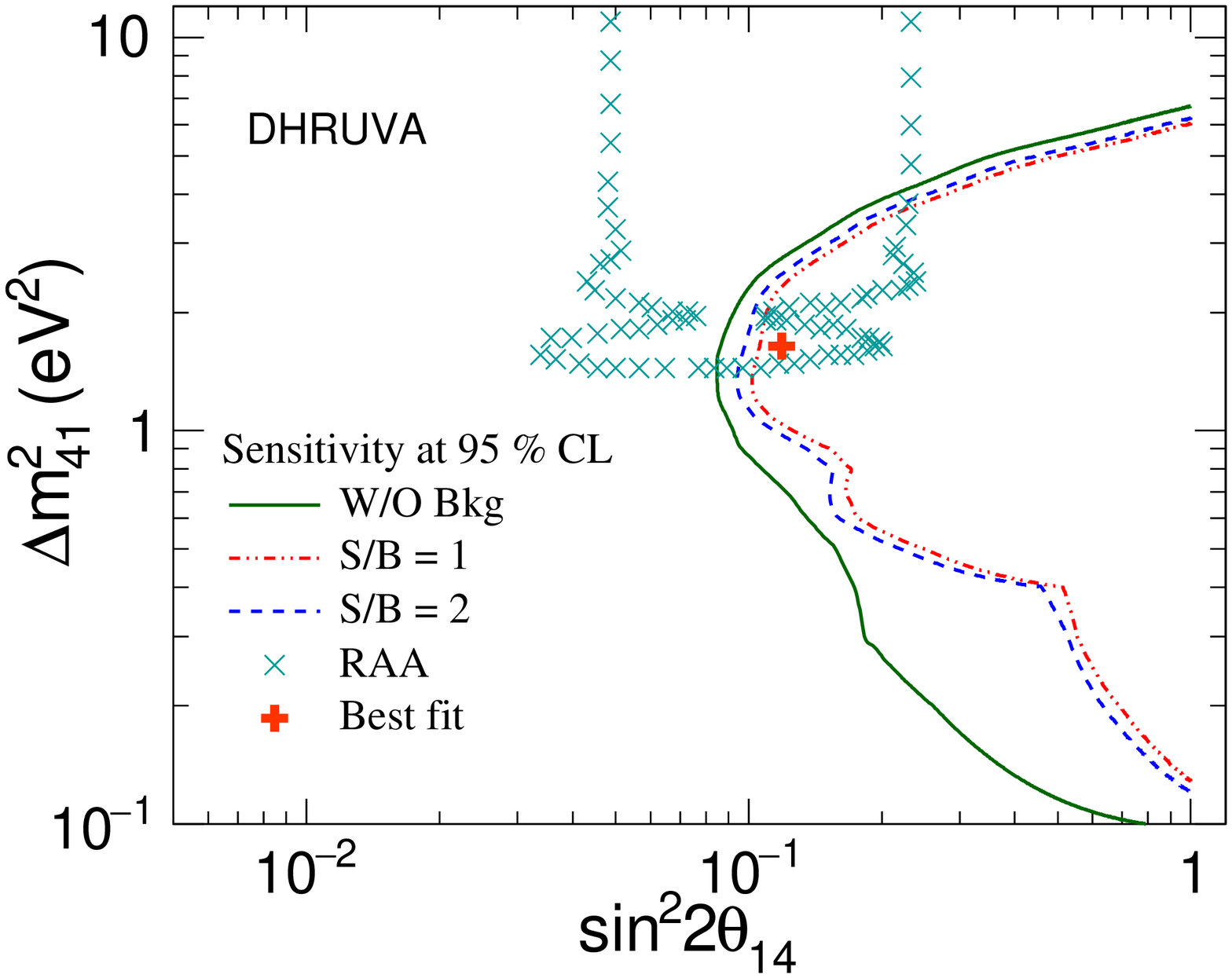}
\caption{Comparison of the ISMRAN sensitivity to the expected active-sterile 
neutrino mixing with and without inclusion of background for an exposure of 1 
ton-yr. A single detector will be placed at a combination of near and far
distances of 7 m and 9 m, respectively from the DHRUVA reactor.}
\label{fig:bkgtwodet}
\ec
\eef

Figure~\ref{fig:global} shows the comparison between the expected sensitivity 
of the ISMRAN and other experimental observations such as 
PROSPECT~\cite{Ashenfelter:2018iov}, DANSS~\cite{Alekseev:2018efk}, 
STEREO~\cite{AlmazanMolina:2019qul}, and Neutrino-4~\cite{Serebrov:2018vdw} 
groups. As far as ISMRAN setup is concerned, the same detector will be placed at 
a combination of near and far positions from a given reactor core for a total 
exposure of 1 ton-yr. The sensitivity of the ISMRAN setup can be comparable with 
DANSS at $\Delta m^2_{41}$ = 0.1 $eV^2$ if it will be placed at the near and far 
positions of 15 m and 17 m from the PFBR reactor core. It can be seen that at 
higher $\Delta m^2_{41} (>$ 1.0 $eV^2$), the sensitivity of the ISMRAN is comparable 
with the results from DANSS experiment, although Neutrino-4 experiment has better 
sensitivity as compared to all other mentioned observations. At low
$\Delta m^2_{41}$($<$ 0.2 MeV$^2$), ISMRAN sensitivity is comparable with
Neutrino-4 observation if it will be placed at DHRUVA reactor facility. 
The present study shows that significant portions of the 
allowed RAA region can be excluded when the detector is placed at the 
near and far positions of 7 m and 9 m from the DHRUVA reactor core.
\bef[ht!]
\bc
\includegraphics[width=0.47\textwidth]{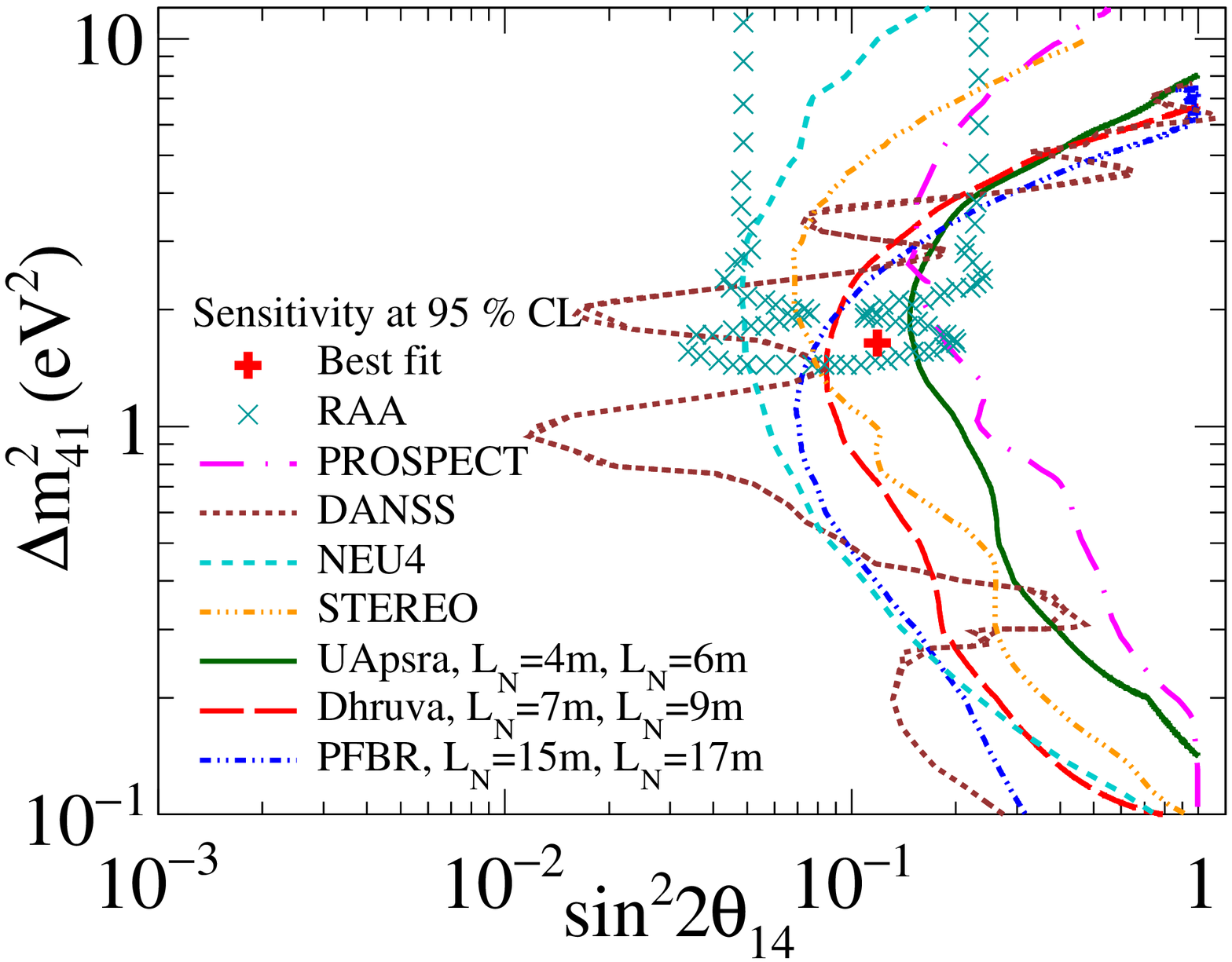}
\caption{The expected active-sterile neutrino mixing sensitivity of the ISMRAN setup 
in the sin$^{2}2\theta_{14} - \Delta m^{2}_{41}$ plane at 95$\%$ C.L. for an 
exposure of 1 ton-yr compared with other experimental observations. The detector is 
placed at a combination of near and far distances from the reactor different cores.
NEU4 is the result from Neutrino-4 experiment}.
\label{fig:global}
\ec
\eef
\section{SUMMARY}
\label{sec:summary}
The existence of sterile neutrinos as the possible origin of the RAA and the origin 
of the 5 MeV bump in the $\overline{\nu}_e$ energy spectra is being explored by 
several SBL experiments using reactor antineutrinos as a source. In the present 
study, we have investigated the potential of the upcoming ISMRAN experimental setup 
for finding out the possible presence of active to sterile neutrino oscillations. The 
analysis is performed for an exposure of 1 ton-year using $\overline{\nu}_e$s 
produced from the U-Apsra, DHRUVA, and PFBR reactor facilities, India. The 
oscillation parameters (sin$^2 2\theta_{14}$, $\Delta m^{2}_{41}$) are constrained by 
considering a single detector which will be placed at either a fixed position  
or combining the observations taken at two different positions with respect to 
the reactor core. The main advantage of putting the same detector at 
different distances is to cancel the systematic uncertainties related to 
the reactor and detector. It is found that the ISMRAN setup can exclude a 
small portion of the favored non-zero active-sterile mixing parameters region 
obtained from RAA with a single detector placed at a fixed position. A 
combination of detector positions can have better sensitivity in excluding 
the RAA region as well as the best fit point compared to a detector placed at 
fixed location, for a given exposure. One of the possible combinations of near 
and far positions for the detector is 7 m and 9 m from the DHRUVA reactor core. 
This gave a better constraint of the RAA compared to other combinations. At lower 
$\Delta m^{2}_{41}$ ($\sim$ 0.1 $eV^2$), the detector can have better sensitivity to 
the active-sterile oscillation parameters, if we place it at the PFBR reactor 
facility with a combination of near and far positions of 15 m and 17 m, 
respectively, from the core, due to its relatively compact core size and large 
thermal power. The sensitivity of the detector could be improved further with 
increased statistics by placing the target volume closer to the reactor and improving 
the signal to background ratios.
\section*{ACKNOWLEDGMENTS}
We thank Anushree Ghosh for helpful suggestions and useful discussions.
We also thank ISMRAN group members for useful discussion.
We thank P. Garg and K. Finnelli for critically reading the manuscript.
  We would also like to take this opportunity to thank the anonymous referee
  for the valuable comments which helped us to improve the scope and
  content of the paper.

\end{document}